\documentclass{article}
\usepackage[top=30mm, bottom=30mm, left=25mm, right=30mm]{geometry}
\usepackage{hyperref}
\usepackage{amsmath}
\usepackage{stmaryrd}
\usepackage{graphicx}
\usepackage{float}
\usepackage{cite}
\usepackage[T1]{fontenc}
\usepackage{array}
\usepackage{textcomp}
\usepackage{setspace}
\usepackage{esint}
\usepackage{amstext}
\usepackage[affil-it]{authblk}
\providecommand{\tabularnewline}{\\}

\begin{document}

\title{Decoupling the NLO coupled QED$\otimes$QCD, DGLAP evolution equations, using Laplace transform method}

\author{Marzieh Mottaghizadeh, Parvin Eslami\thanks{Corresponding author: eslami@um.ac.ir} and Fatemeh Taghavi Shahri}

\affil{Department of Physics, Ferdowsi University of Mashhad,
Mashhad, Iran}\maketitle

\begin{abstract}
We analytically solved the QED{\normalsize{}$\otimes$}QCD coupled
DGLAP evolution equations at leading order (LO) quantum electrodynamics
(QED) and next to leading order (NLO) quantum chromodynamics (QCD)
approximations, using the Laplace transform method and then computed
the proton structure function in terms of the unpolarized parton distributions
functions. Our analyitical solutions for parton densities are in good
agreement with those from CT14QED ($1.295^{2}<Q^{2}<10^{10}$) (Phys.
Rev. D 93, 114015 (2016)) global parameterizations and APFEL (A PDF
Evolution Library) ($2<Q^{2}<10^{8}$) (Computer Physics Communications
185, 1647-1668 (2014)). We also compared the proton structure function,
$F_{2}^{p}(x,Q^{2})$, with experimental data released by the ZEUS
and H1 collaborations at HERA. There is a nice agreement between them
in the range of low and high x and $Q^{2}$.

\end{abstract}

\section{Introduction}\label{sec:intro}

Accurate determination of the parton distribution function (PDF) inside
proton is an essential part of analyzing data in deep-inelastic scattering
(DIS) processes.

Precise measurements from high energy hadron colliders such as Tevatron
and Large Hadron Collider (LHC) require the inclusion of higher order
effects in proton-proton scattering. It seems that the photon-induced
Drell-Yan (DY) process such as $\gamma\gamma\rightarrow l^{+}l^{-}$has
a significant contribution ($\sim10\%$ ) to the dilepton invariant
mass distribution. Recent results from high mass DY production in
ATLAS ~\cite{Aad:2013iua}  showed that contribution of
photon distribution inside proton has the same importance as the other
different PDFs set. To calculate the cross section of such DY process,
one needs to know the photon distribution function inside proton,
$\gamma(x,Q^{2})$. Furthermore, because the LHC is really a $\gamma\gamma$
collider at very high energy, then the determination of photon distribution
function inside proton may be an important issue.

There are a few studies about adding the QED corrections to the global
parameterizations of PDFs which are based on QCD calculations. The
first one have been done by the MRST group ~\cite{Martin:1998sq,Martin:2004dh} 
and the other analysis are newly released by NNPDF collaboration ~\cite{Bertone:2013vaa,Ball:2013hta}
and CT14QED group ~\cite{Schmidt:2015zda}.

Here, we will study the analytical solutions for DGLAP evolution equations
to obtain the parton distribution functions at NLO QCD and LO QED
approximations based on the Laplace transform technique which has
introduced by Block et al ~\cite{Block:2007pg,Block:2009en,Block:2010du,Block:2010fk,Block:2010ti,Block:2011px,Block:2011xb}. 

Recently, H. Khanpour et al. ~\cite{Khanpour:2016uxh}
calculate the proton structure function and parton distribution functions
using the Laplace transform technique at NLO in QCD without QED corrections.
They consider the initial value of parton distribution functions from
KKT12~\cite{Khanpour:2012tk} and GJR08 ~\cite{Gluck:2008gs}
codes at $Q_{0}^{2}=2GeV^{2}$. 

The Laplace transform method has an ability that the analytical solutions
for the QED$\otimes$QCD parton distribution functions are obtained
more strictly by using the related kernels and the calculations can
be control well. Following our recent works ~\cite{TaghaviShahri:2010be,AtashbarTehrani:2013qea,Zarei:2015jvh}
on analytical solution of DGLAP evolution equations based on the Laplace
transform, we have used the same method to solve the QED$\otimes$QCD
DGLAP evolution equations. 

The paper is organized as follows. In Section \ref{sec:2},
we review the QED$\otimes$QCD coupled DGLAP evolution equations.
In Section \ref{sec:2}, we bring out the analytically solutions
for the DGLAP evolution equations to calculate the PDFs inside proton
based on the Laplace transform. Section \ref{sec:results} is devoted
to the results for different kind of the parton distribution functions
and also the proton structure function. To be sure about correctness
of our analytical solutions, the final results were cross-checked
with the same results from APFEL (A PDF Evolution Library) program
and also with newly released CT14QED code, we are selected our initial
inputs from CT14QED code at $Q_{0}=1.295\,GeV$. Finally we give our
summary and conclusions in Section\ref{sec:con}. 

\section{REVIEW OF THE QED$\otimes$QCD DGLAP EVOLUTION EQUATIONS}\label{sec:2}

The QED$\otimes$QCD DGLAP evolution equations for the quark, gloun
and the photon parton densities can be written as ~\cite{Balitsky:1978ic,Altarelli:1977zs,Carrazza:2015dea}: 
\begin{eqnarray}
\frac{\partial{{q}_{i}}}{\partial\ln{{Q}^{2}}}&=&\sum\limits _{j=1}^{{{n}_{f}}}{P_{{{q}_{i}}{{q}_{j}}}^ {}(x)\otimes{{q}_{j}}+}\sum\limits _{j=1}^{{{n}_{f}}}{P_{{{q}_{i}}{{\bar{q}}_{j}}}^ {}(x)\otimes{{\bar{q}}_{j}}}
+{{P}_{{{q}_{i}}g}}\otimes g+{{P}_{{{q}_{i}}\gamma}}\otimes\gamma
\nonumber\\
\frac{\partial{{\bar{q}}_{i}}}{\partial\ln{{Q}^{2}}}&=&\sum\limits _{j=1}^{{{n}_{f}}}{P_{{{\bar{q}}_{i}}{{q}_{j}}}^ {}(x)\otimes{{q}_{j}}+}\sum\limits _{j=1}^{{{n}_{f}}}{P_{{{\bar{q}}_{i}}{{\bar{q}}_{j}}}^ {}(x)\otimes{{\bar{q}}_{j}}}
+{{P}_{{{\bar{q}}_{i}}g}}\otimes g+{{P}_{{{\bar{q}}_{i}}\gamma}}\otimes\gamma
\nonumber\\
\frac{\partial g}{\partial\ln{{Q}^{2}}}&=&\sum\limits _{j=1}^{{{n}_{f}}}{P_{g{{q}_{j}}}^ {}(x)\otimes{{q}_{j}}+}\sum\limits _{j=1}^{{{n}_{f}}}{P_{g{{\bar{q}}_{j}}}^ {}(x)\otimes{{\bar{q}}_{j}}}
+{{P}_{gg}}\otimes g
\nonumber\\
\frac{\partial\gamma}{\partial\ln{{Q}^{2}}}&=&\sum\limits _{j=1}^{{{n}_{f}}}{P_{\gamma{{q}_{j}}}^ {}(x)\otimes{{q}_{j}}+}\sum\limits _{j=1}^{{{n}_{f}}}{P_{\gamma{{\bar{q}}_{j}}}^ {}(x)\otimes{{\bar{q}}_{j}}}
+{{P}_{\gamma\gamma}}\otimes\gamma\label{eq1}
\end{eqnarray}

Where $q_{i}(x,Q^{2})$, $\bar{q}_{i}(x,Q^{2})$, $g(x,Q^{2})$ and
$\gamma(x,Q^{2})$ are the i-th quark , i-th antiquark, the gloun
and the photon distribution functions, respectively. The $\otimes$
symbol refers to the convolution integral and the splitting functions
in the right-hand side of Eq.~\ref{eq1}can be written as, 

\begin{eqnarray}
P_{{{q}_{i}}{{\bar{q}}_{j}}}^ {}&=& P_{{{\bar{q}}_{i}}{{q}_{j}}}^ {}=a_{s}^{2}({{\delta}_{ij}}\frac{P_{+}^{(1)}-P_{-}^{(1)}}{2}+\frac{P_{qq}^{(1)}-P_{+}^{(1)}}{2{{n}_{f}}})
\nonumber\\
P_{{{q}_{i}}{{q}_{j}}}^ {} & = & P_{{{\bar{q}}_{i}}{{\bar{q}}_{j}}}^ {}={{a}_{s}}{{\delta}_{ij}}\tilde{P}_{qq}^{(0)}+ a_{s}^{2}({{\delta}_{ij}}\frac{P_{+}^{(1)}+P_{-}^{(1)}}{2}+\frac{P_{qq}^{(1)}-P_{+}^{(1)}}{2{{n}_{f}}})+ a({{\delta}_{ij}}{{e}_{i}}{{e}_{j}})\tilde{P}_{qq}^{(0)}
\nonumber\\
P_{g{{q}_{i}}}^ {}&=&{{P}_{g{{\bar{q}}_{i}}}}={{a}_{s}}P_{gq}^{(0)}+a_{s}^{2}P_{gq}^{(1)}
\nonumber\\
P_{gg}^ {}&=& {{a}_{s}}P_{gg}^{(0)}+a_{s}^{2}P_{gg}^{(1)}
\nonumber\\
P_{\gamma{{q}_{i}}}^ {}&=& P_{\gamma{{\bar{q}}_{i}}}^ {}=ae_{i}^{2}P_{\gamma q}^{(0)}
\nonumber\\
P_{\gamma\gamma}&=& aP_{\gamma\gamma}^{(0)}
\nonumber\\
P_{{{q}_{i}}\gamma}&=& P_{{{\bar{q}}_{i}}\gamma}=ae_{i}^{2}\frac{P_{q\gamma}^{(0)}}{2{{n}_{f}}}\label{eq2}
\end{eqnarray}
The running strong coupling $a_{s}=\alpha_{s}/2\pi$ is determined
by

\begin{eqnarray}
a_{s}(Q^{2}) & =\frac{1}{\beta_{0}Log(\frac{Q^{2}}{\Lambda_{QCD}^{2}})}(1-\frac{\beta_{1}}{\beta_{0}^{2}}\frac{Log(Log(\frac{Q^{2}}{\Lambda_{QCD}^{2}}))}{Log(\frac{Q^{2}}{\Lambda_{QCD}^{2}})}) & ,\label{eq3}
\end{eqnarray}

and the electromagnetic coupling constant in the recent studies ~\cite{Martin:2004dh}
have been considered $\alpha=1/137,$ but here we give $a=\alpha/2\pi$
as follows:

\begin{eqnarray}
a(Q^{2}) & =\frac{a(\mu^{2})}{1-\frac{38}{9}a(\mu^{2})Log(\frac{Q^{2}}{\mu^{2}})} & ,\label{eq4}
\end{eqnarray}

where $\beta_{0}=\frac{1}{3}(33-2n_{f})$ and $\beta_{1}=102-\frac{38}{3}n_{f}$.
For $n_{f}=5$, we get $\Lambda_{QCD}=0.22$. We suppose $\mu=1.777GeV$
then $a(\mu^{2})=\frac{1}{2\pi}\frac{1}{133.4}$ ~\cite{Deur:2016tte}. 

The LO splitting functions are given by ~\cite{Altarelli:1977zs}
\begin{eqnarray}
P_{qq}^{(0)}(x)&=&\frac{4}{3}(\frac{1+{{x}^{2}}}{{{(1-x)}_{+}}}+\frac{3}{2}\delta(1-x))
\nonumber\\
\tilde{P}_{qq}^{(0)}(x)&=&\frac{3}{4}P_{gq}^{(0)}(x)
\nonumber\\
P_{qg}^{(0)}(x)&=&{{n}_{f}}({{x}^{2}}+{{(1-x)}^{2}})
\nonumber\\
P_{q\gamma}^{(0)}(x)&=& 2P_{qg}^{(0)}(x)
\nonumber\\
P_{\gamma q}^{(0)}(x)&=&\frac{4}{3}\left[\frac{1+{{(1-x)}^{2}}}{x}\right]
\nonumber\\
P_{\gamma q}^{(0)}(x)&=&\frac{3}{4}P_{qq}^{(0)}(x)
\nonumber\\
P_{gg}^{(0)}(x)&=& 6(\frac{x}{{{(1-x)}_{+}}}+\frac{1-x}{x}+x(1-x)
+(\frac{11}{12}-\frac{{{n}_{f}}}{18})\delta(1-x))
\nonumber\\
\tilde{P}_{\gamma\gamma}^{(0)}(x)&=&-\frac{2}{3}\sum\limits _{i=1}^{{{n}_{f}}}{e_{i}^{2}\delta(1-x)}\label{eq5}
\end{eqnarray}
The $P_{qq}^{(1)}$, $P_{qg}^{(1)}$, $P_{gq}^{(1)}$ and $P_{gg}^{(1)}$used
in Eq. ~\ref{eq2} are the NLO singlet
splitting functions, $P_{+}^{(1)}$and $P_{-}^{(1)}$are the NLO non-singlet
splitting functions that can be found in Refs. ~\cite{Furmanski:1980cm,Curci:1980uw}. 

For the coupled approach we utilize a PDF basis for the QED$\otimes$QCD
DGLAP evolution equations, defined by the following singlet and non-singlet
PDF combinations ~\cite{Roth:2004ti},

\begin{center}
\begin{equation}
q^{SG}:\left(\begin{array}{c}
{{f}_{1}}=\Delta=\\
u+\bar{u}+c+\bar{c}-d-\bar{d}-s-\bar{s}-b-\bar{b}\\
{{f}_{2}}=\Sigma=\\
u+\bar{u}+c+\bar{c}+d+\bar{d}+s+\bar{s}+b+\bar{b}\\
{{f}_{3}}=g\\
{{f}_{4}}=\gamma
\end{array}\right)\label{eq6}
\end{equation}
\par\end{center}

\begin{center}
\begin{equation}
q^{NS}:\left(\begin{array}{c}
{{f}_{5}}={{d}_{v}}=d-\bar{d}\\
{{f}_{6}}={{u}_{v}}=u-\bar{u}\\
{{f}_{7}}={{\Delta}_{ds}}=d+\bar{d}-s-\bar{s}\\
{{f}_{8}}={{\Delta}_{uc}}=u+\bar{u}-c-\bar{c}\\
{{f}_{9}}={{\Delta}_{sb}}=s+\bar{s}-b-\bar{b}
\end{array}\right)\label{eq7}
\end{equation}
\par\end{center}

We have found the singlet PDFs evolve as:

\begin{eqnarray}
\frac{\partial}{\partial lnQ^{2}}\left(\begin{array}{c}
f_{1}\\
f_{2}\\
f_{3}\\
f_{4}
\end{array}\right)=
\left(\begin{array}{cccc}
P_{11} & P_{12} & P_{13} & P_{14}\\
P_{21} & P_{22} & P_{23} & P_{24}\\
P_{31} & P_{32} & P_{33} & P_{34}\\
P_{41} & P_{42} & P_{43} & P_{44}
\end{array}\right)\otimes\left(\begin{array}{c}
f_{1}\\
f_{2}\\
f_{3}\\
f_{4}
\end{array}\right)\label{eq8}
\end{eqnarray}

and the non-singlet PDFs, obey the evolution equations such as:

\begin{equation}
\frac{\partial{{f}_{i}}}{\partial\ln{{Q}^{2}}}={{P}_{ii}}\otimes{{f}_{i}}\qquad i=5,\ldots,9\label{eq9}
\end{equation}

In the equations ~\ref{eq8}and ~\ref{eq9}
the new splitting functions are calculated as

\begin{eqnarray}
{{P}_{11}}&=&{{a}_{s}}P_{qq}^{(0)}+a_{s}^{2}P_{+}^{(1)}+\frac{e_{u}^{2}+e_{d}^{2}}{2}a\tilde{P}_{qq}^{(0)}
\nonumber\\
{{P}_{12}}&=&\frac{{{n}_{u}}-{{n}_{d}}}{{{n}_{f}}}a_{s}^{2}(P_{qq}^{(1)}-P_{+}^{(1)})+\frac{e_{u}^{2}-e_{d}^{2}}{2}a\tilde{P}_{qq}^{(0)}
\nonumber\\
{{P}_{13}}&=&\frac{{{n}_{u}}-{{n}_{d}}}{{{n}_{f}}}({{a}_{s}}P_{qg}^{(0)}+a_{s}^{2}P_{qg}^{(1)})
\nonumber\\
{{P}_{14}}&=&\frac{{{n}_{u}}e_{u}^{2}-{{n}_{d}}e_{d}^{2}}{{{n}_{f}}}aP_{q\gamma}^{(0)}
\nonumber\\
{{P}_{21}}&=&\frac{e_{u}^{2}-e_{d}^{2}}{2}a\tilde{P}_{qq}^{(0)}
\nonumber\\
{{P}_{22}}&=&{{a}_{s}}P_{qq}^{(0)}+a_{s}^{2}P_{qq}^{(1)}+\frac{e_{u}^{2}+e_{d}^{2}}{2}a\tilde{P}_{qq}^{(0)}
\nonumber\\
{{P}_{23}}&=&{{a}_{s}}P_{qg}^{(0)}+a_{s}^{2}P_{qg}^{(1)}
\nonumber\\
{{P}_{24}}&=&\frac{{{n}_{u}}e_{u}^{2}+{{n}_{d}}e_{d}^{2}}{{{n}_{f}}}aP_{q\gamma}^{(0)}
\nonumber\\
{{P}_{31}}&=& 0
\nonumber\\
{{P}_{32}}&=&{{a}_{s}}P_{gq}^{(0)}+a_{s}^{2}P_{gq}^{(1)}
\nonumber\\
{{P}_{33}}&=&{{a}_{s}}P_{gg}^{(0)}+a_{s}^{2}P_{gg}^{(1)}
\nonumber\\
{{P}_{34}}&=& 0
\nonumber\\
{{P}_{41}}&=&\frac{e_{u}^{2}-e_{d}^{2}}{2}aP_{\gamma q}^{(0)}
\nonumber\\
{{P}_{42}}&=&\frac{e_{u}^{2}+e_{d}^{2}}{2}aP_{\gamma q}^{(0)}
\nonumber\\
{{P}_{43}}&=& 0
\nonumber\\
{{P}_{44}}&=&{{a}_{{}}}P_{\gamma\gamma}^{(0)}
\nonumber\\
{{P}_{55}}&=&{{a}_{s}}P_{qq}^{(0)}+a_{s}^{2}P_{-}^{(1)}+ae_{d}^{2}\tilde{P}_{qq}^{(0)}
\nonumber\\
{{P}_{66}}&=&{{a}_{s}}P_{qq}^{(0)}+a_{s}^{2}P_{-}^{(1)}+ae_{u}^{2}\tilde{P}_{qq}^{(0)}
\nonumber\\
{{P}_{77}}&=&{{P}_{99}}={{a}_{s}}P_{qq}^{(0)}+a_{s}^{2}P_{+}^{(1)}+ae_{d}^{2}\tilde{P}_{qq}^{(0)}
\nonumber\\
{{P}_{88}}&=&{{a}_{s}}P_{qq}^{(0)}+a_{s}^{2}P_{+}^{(1)}+ae_{u}^{2}\tilde{P}_{qq}^{(0)}\label{eq10}
\end{eqnarray}
where $n_{u}$ and $n_{d}$ are the number of up- and down-type active
quark flavors, respectively, and $n_{f}=n_{u}+n_{d}$. In the next
section, we try to solve the above equations with Laplace transform
method.

\section{THE ANALYTICAL SOLUTIONS OF THE QED$\otimes$QCD
DGLAP EVOLUTION EQUATIONS}\label{sec:3}

Now, we are in a position to briefly review the method of extracting
the PDFs via the analytical solutions of DGLAP evolution equations
using the Laplace transform technique. Block et al. , in Ref. ~\cite{Block:2010du},
showed that, using the Laplace transform, one can solve the DGLAP
evolution equations directly and extract unpolarized parton distribution
functions. We will give the details here and review the method for
extracting the unpolarized parton distribution functions of QED$\otimes$QCD
coupled DGLAP equations at LO QED and NLO QCD approximations. By introducing
the variables $\nu\equiv\ln(1/x)$ and $\tau({{Q}^{2}},Q_{0}^{2})\equiv\frac{1}{2\pi}\int_{Q_{0}^{2}}^{{{Q}^{2}}}{{{\alpha}_{s}}({{{Q}'}^{2}})}d\ln{{{Q}'}^{2}}$
into the coupled DGLAP equations, one can turn them into coupled convolution
equations in $\nu$ and $\tau$ spaces. We use two Laplace transforms
from $\nu$ and $\tau$ spaces to s and U spaces, respectively, then
the DGLAP equations can be solved iteratively by a set of convolution
integrals which are dependent on unpolarized parton distribution functions
at an initial input scale of $Q_{0}^{2}$.

In the following subsections, 3.1 and 3.2, we present solutions of
equations~\ref{eq8} and ~\ref{eq9}
separately. 

\subsection{The Singlet Solution}

By considering the variable changes $\nu\equiv\ln(1/x)$ and $w\equiv ln(1/z)$,
one can rewrite the equations~\ref{eq8}
in terms of the convolution integrals, as

\begin{eqnarray}
&\frac{\partial\hat{F}_{i}}{\partial\tau}(v,\tau)=\intop_{0}^{v}\sum\limits _{j=1}^{4}(\hat{K}_{ij}^{LO,QCD}(v-w)
+\frac{\alpha}{{{\alpha}_{s}}}\hat{K}_{ij}^{LO,QED}(v-w)+\frac{\alpha_{s}}{2\pi}\hat{K}_{ij}^{NLO,QCD}(v-w))
\hat{F}_{j}(w,\tau)dw\qquad 
\nonumber\\&
i=1,\ldots,4\label{eq11}
\end{eqnarray}

Note that we have used the notation $\hat{F}_{i}(v,\tau)\equiv F_{i}(e^{-v},\tau)$.
The above convolution integrals show that $\hat{K}_{ij}(v)\equiv e^{-v}P_{ij}(e^{-v})$.

Using this fact that the Laplace transform of a convolution simply
is the ordinary product of the Laplace transform of the factors, the
Laplace transform from $\nu$ space to s space convert Eq.~\ref{eq11}to ordinary
first order differential equations

\begin{eqnarray}
\frac{\partial{{f}_{i}}}{\partial\tau}(s,\tau)=\sum\limits _{j=1}^{4}(\Phi_{ij}^{LO,QCD}+\frac{\alpha}{{{\alpha}_{s}}}\Phi_{ij}^{LO,QED}
+\frac{\alpha_{s}}{2\pi}\varPhi_{ij}^{NLO,QCD}){{f}_{j}}(s,\tau)\qquad i=1,\ldots,4\label{eq12}
\end{eqnarray}

Here we intend to extend our calculations to the NLO approximation
for the $\Delta$, $\Sigma$ , gluon and photon sectors of unpolarized
parton distributions. In this case, to decouple and to solve DGLAP
evolution equations ~\ref{eq12} we need an extra Laplace transformation from $\tau$ space to U space.
In the rest of the calculation, the $\alpha_{s}(\tau)/2\pi$ and $\alpha(\tau)/\alpha_{s}(\tau)$
are replaced for brevity by $a^{QCD}(\tau)$ and $a^{QED}(\tau)$
, respectively. Therefore the solutions of the first order differential
equations in Eq.~\ref{eq12} can be converted to,

\begin{eqnarray}
&
U{{F}_{i}}(s,U)-{{f}_{i0}}(s)\text{ }=\sum\limits _{j=1}^{4}\Phi_{ij}^{LO,QCD}(s)\text{ }{{F}_{j}}(s,U)
+\Phi_{ij}^{LO,QED}(s)\text{ }L[a^{QED}(\tau){{f}_{j}}(s,\tau);U]
\nonumber\\&
+\Phi_{ij}^{NLO,QCD}(s)\text{ }L[a^{QCD}(\tau){{f}_{j}}(s,\tau);U]      
i=1,\ldots,4\label{eq13}
\end{eqnarray}

To simplify the NLO calculations we use two excellent approximation
relations $a^{QCD}(\tau)={{a}_{0}}+{{a}_{1}}{{e}^{-{{b}_{1}}\tau}}$
, where $a_{0}=0.003$, $a_{1}=0.05$ and $b_{1}=4.9$ and also $a^{QED}(\tau)=-{{\tilde{a}}_{0}}+{{\tilde{a}}_{1}}{{e}^{-{{\tilde{b}}_{1}}\tau}}$,
where $\tilde{a}_{0}=-0.0036$ , $\tilde{a}_{1}=0.025$ and $\tilde{b}_{1}=-3.9$
for $M_{b}^{2}<Q^{2}\le10^{8}GeV^{2}$.

Therefore, we write expressions $L[a^{QCD}(\tau){{f}_{j}}(s,\tau);U]$
and $L[a^{QED}(\tau){{f}_{j}}(s,\tau);U]$ needed in Eq.~\ref{eq13} as

\begin{eqnarray}
L[a^{QCD}(\tau){{f}_{j}}(s,\tau);U]=\sum\limits _{j=0}^{1}a_{j}F(s,U+b_{j}),\nonumber \\
L[a^{QED}(\tau){{f}_{j}}(s,\tau);U]=\sum\limits _{j=0}^{1}\tilde{a}_{j}F(s,U+\tilde{b}_{j}),\label{eq14}\\
b_{0}=0\,and\,\tilde{b}_{0}=0.\nonumber 
\end{eqnarray}

After introducing the simplifying notations for the splitting functions,
we will have

\begin{eqnarray}
{{\Phi}_{ij}}(s)=\Phi_{ij}^{LO,QCD}(s)+{{\tilde{a}}_{0}}\Phi_{ij}^{LO,QED}(s)+{{a}_{0}}\Phi_{ij}^{NLO,QCD}(s)\qquad i,j=1,\ldots,4\label{eq15}
\end{eqnarray}
Therefore, the solutions of the first order differential equations
in Eq.~\ref{eq13}
can be changed to,

\begin{eqnarray}
[U-{{\Phi}_{ii}})]{{\tilde{F}}_{i}}(s,U)-\sum\limits _{j=2}^{4}\Phi_{ij}^ {}{{\tilde{F}}_{j}}(s,U)={{f}_{i0}}(s)+{{\tilde{a}}_{1}}[\sum\limits _{j=i}^{4}\Phi_{ji}^{LO,QED}{{F}_{j}}(s,U+{{\tilde{b}}_{1}})]
\nonumber\\
+{{a}_{1}}[\sum\limits _{j=i}^{4}\Phi_{ji}^{NLO,QCD}{{F}_{j}}(s,U+{{b}_{1}})]\qquad i,j=1,\ldots,4\label{eq16}
\end{eqnarray}

The complete solutions of Eq. ~\ref{eq16}
can be obtained via iteration processes. The iteration can be continued
to any required order but we will restrict our selves in which we
get to a sufficient convergence of the solutions. Our results show
that the second order of iteration are sufficient to get a reasonable
convergence. Using the first inverse Laplace transform technique~\cite{Block:2011px} from U space to $\tau$ space, we can obtain the following expression for the distributions,

\begin{equation}
{{f}_{i}}(s,\tau)=\sum\limits _{j=1}^{4}{{{k}_{ij}}({{a}_{1}},{{b}_{1}},s,\tau){{f}_{j0}}(s)}\label{eq17}
\end{equation}

With the initial input functions for $\Sigma$ , $\Delta$, gloun
and photon sectors of distributions, which are denoted by ${{f}_{10}}(s)$,
${{f}_{20}}(s)$, ${{f}_{30}}(s)$ and ${{f}_{40}}(s)$, respectively.
By the second inverse Laplace transform from s space to $\nu\equiv\ln(1/x)$
space, we get parton distribution functions in the usual x space. 

\subsection{The Non-Singlet Solution}

We perform here the non-singlet solutions of the QED$\otimes$QCD
DGLAP evolution equation, ~\ref{eq9},
using the Laplace transform technique at LO QED and NLO QCD approximations.
For the non-singlet distributions $\hat{F}_{i}(\nu,\tau)$, after
changing to the variable $v\equiv ln(1/x)$ and the variable $\tau$, we can schematically write the equation~\ref{eq9}as

\begin{eqnarray}
\frac{\partial{{\hat{F}}_{i}}}{\partial\tau}(\nu,\tau)=\int_{0}^{\nu}{{\hat{F}}_{i}}(w,\tau){{e}^{-(\nu-w)}}
{P}(\nu-w)dw\text{ }\qquad i=5,\ldots,9\label{eq18}
\end{eqnarray}
where

\begin{equation}
\hat{F}_{i}(v,\tau)\equiv F_{i}(e^{-v},\tau)\qquad i=5,\ldots,9\label{eq19}
\end{equation}

Going to Laplace space s, we can obtain first order differential equations
with respect to $\tau$ variable for the non-singlet distributions
$f_{i,ns}(s,\tau)$, whose solutions are,

\begin{equation}
{{f}_{i,\text{ns}}}(s,\tau)={{e}^{\tau{{\Phi}_{ns}}(s)}}{{f}_{i,\text{ns}0}}(s)\;i=5,\ldots,9\text{ }\label{eq20}
\end{equation}

For example, for valence quarks, such as $U_{val}=x(u(x,Q^{2})-\bar{u}(x,Q^{2}))$,
$\varPhi_{ns}(s)$ can be written as

\begin{eqnarray}
{{\Phi}_{ns}}(s)=\Phi_{\text{ns}}^{LO,QCD}+\frac{{{\tau}_{2}}}{\tau}\Phi_{\text{ns}}^{LO,QED}
+\frac{{{\tau}_{3}}}{\tau}\Phi_{\text{ns}}^{NLO,QCD}\text{ }\label{eq21}
\end{eqnarray}

where

\begin{equation}
\Phi_{ns}^{LO,QCD}=L[{{e}^{-v}}P_{qq}^{LO}({{e}^{-v}});s]
\nonumber
\end{equation}

\begin{equation}
\Phi_{ns}^{LO,QED}\text{ }=e_{u}^{2}L[{{e}^{-v}}\tilde{P}_{qq}^{LO}({{e}^{-v}});s]
\nonumber
\end{equation}

\begin{equation}\
\Phi_{ns}^{NLO,QCD}=L[{{e}^{-v}}P_{qq}^{NLO}({{e}^{-v}});s]
\nonumber
\end{equation}
The $\tau_{2}$ and $\tau_{3}$ parameters in Eq.~\ref{eq21}
are defined as

\begin{equation}
\tau_{2}\equiv\frac{1}{2\pi}\int_{0}^{\tau}{\alpha(\tau')}d\ln{{\tau}'}=
\frac{1}{(2\pi)^{2}}\int_{Q_{0}^{2}}^{{{Q}^{2}}}{\alpha(Q'^{2}){{\alpha}_{s}}({{{Q}'}^{2}})}d\ln{{{Q}'}^{2}}
\nonumber
\end{equation}

\begin{equation}
\tau_{3}\equiv\frac{1}{2\pi}\int_{0}^{\tau}{\alpha_{s}(\tau')}d\ln{{\tau}'}=
\frac{1}{(2\pi)^{2}}\int_{Q_{0}^{2}}^{{{Q}^{2}}}{\alpha_{s}^{2}(Q'^{2})}d\ln{{{Q}'}^{2}}
\nonumber
\end{equation}

The $\tau_{2}$ parameter is related to the leading order QED running
coupling constant. The non-singlet solutions, $f_{i}(x,Q^{2})$, can
be obtained using the non-singlet kernel $K_{ns}(v)=L^{-1}[e^{\tau\Phi_{ns}(s)};v]$
in the convolution integral
\begin{equation}
{{\hat{F}}_{ns}}(\nu,\tau)=\int_{0}^{\nu}{{{K}_{ns}}(v-w,\tau){{\hat{F}}_{ns0}}(w)dw\text{ }}\label{eq22}
\end{equation}

Finally, with these two Laplace transforms, the evolution equations
~\ref{eq22} can be solved iteratively
by a set of convolution integrals which are related to the quark distributions
at an initial input scale of $Q_{0}^{2}$ in $(x,Q^{2})$ space.

\subsection{RESULTS AND DISCUSSION}\label{sec:results}

\begin{figure}[tph]
\centering
\includegraphics[scale=0.4]{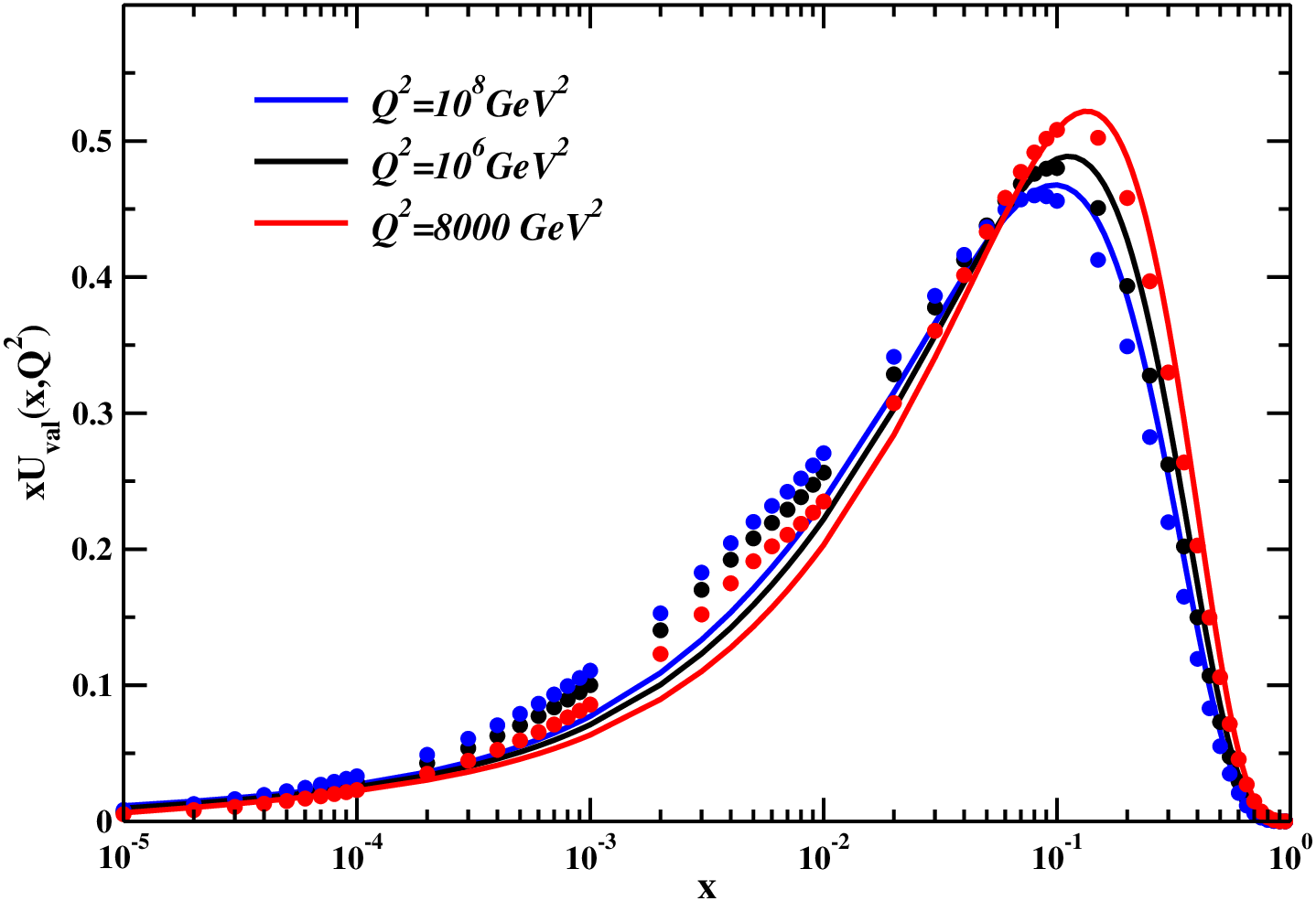}

\centering{}\caption{The $xU_{val}(x,\,Q)$ valance quark distributions in different values
of $Q^{2}$ in comparison with APFEL model. }
\label{fig1}
\end{figure}

\begin{figure}[tph]
\centering
\includegraphics[scale=0.4]{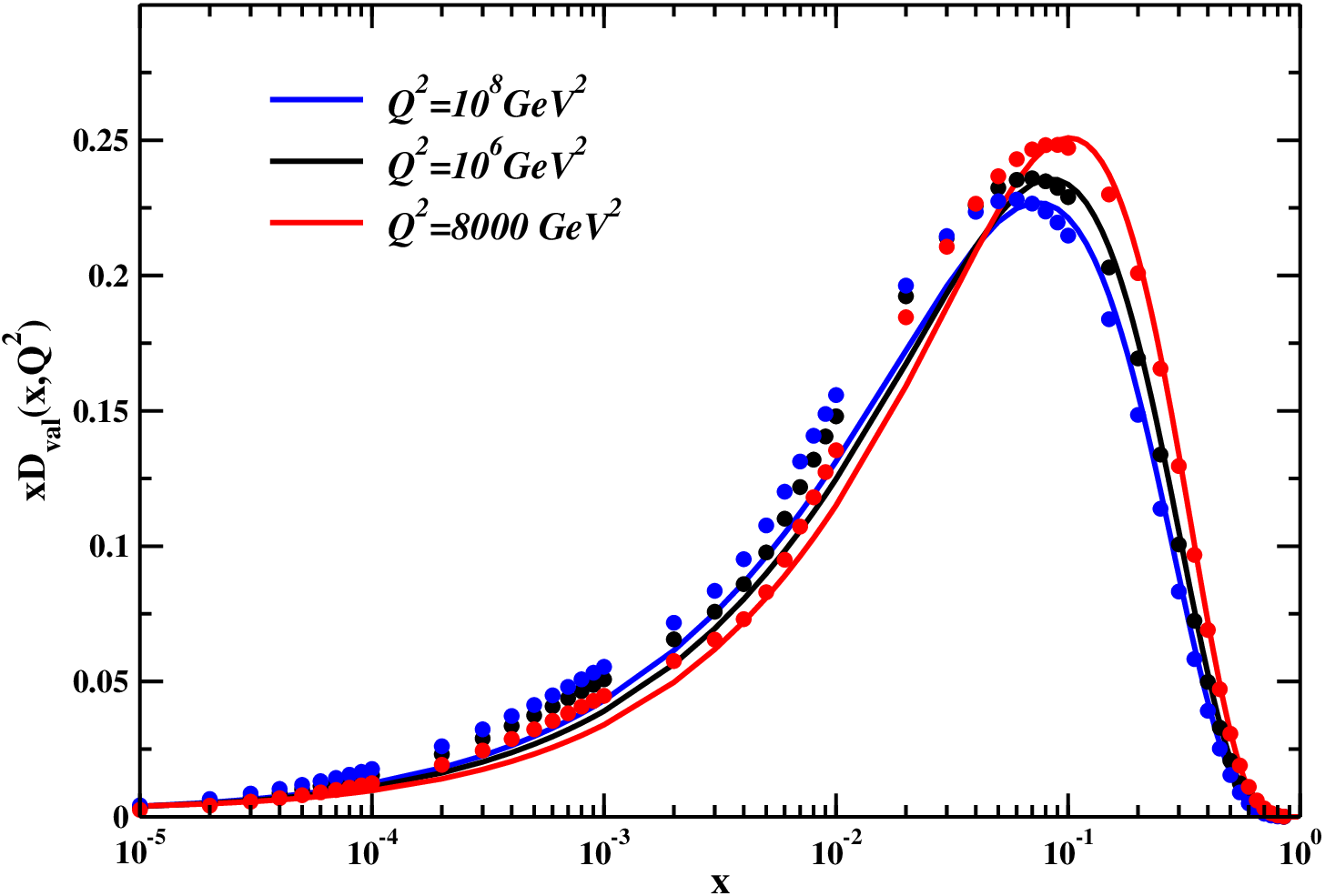}

\caption{The $xD_{val}(x,\,Q^{2})$ valance quark distributions in different
values of $Q^{2}$ in comparison with APFEL model. }

\label{fig2}
\end{figure}

\begin{figure}[ph]
\centering
\includegraphics[scale=0.5]{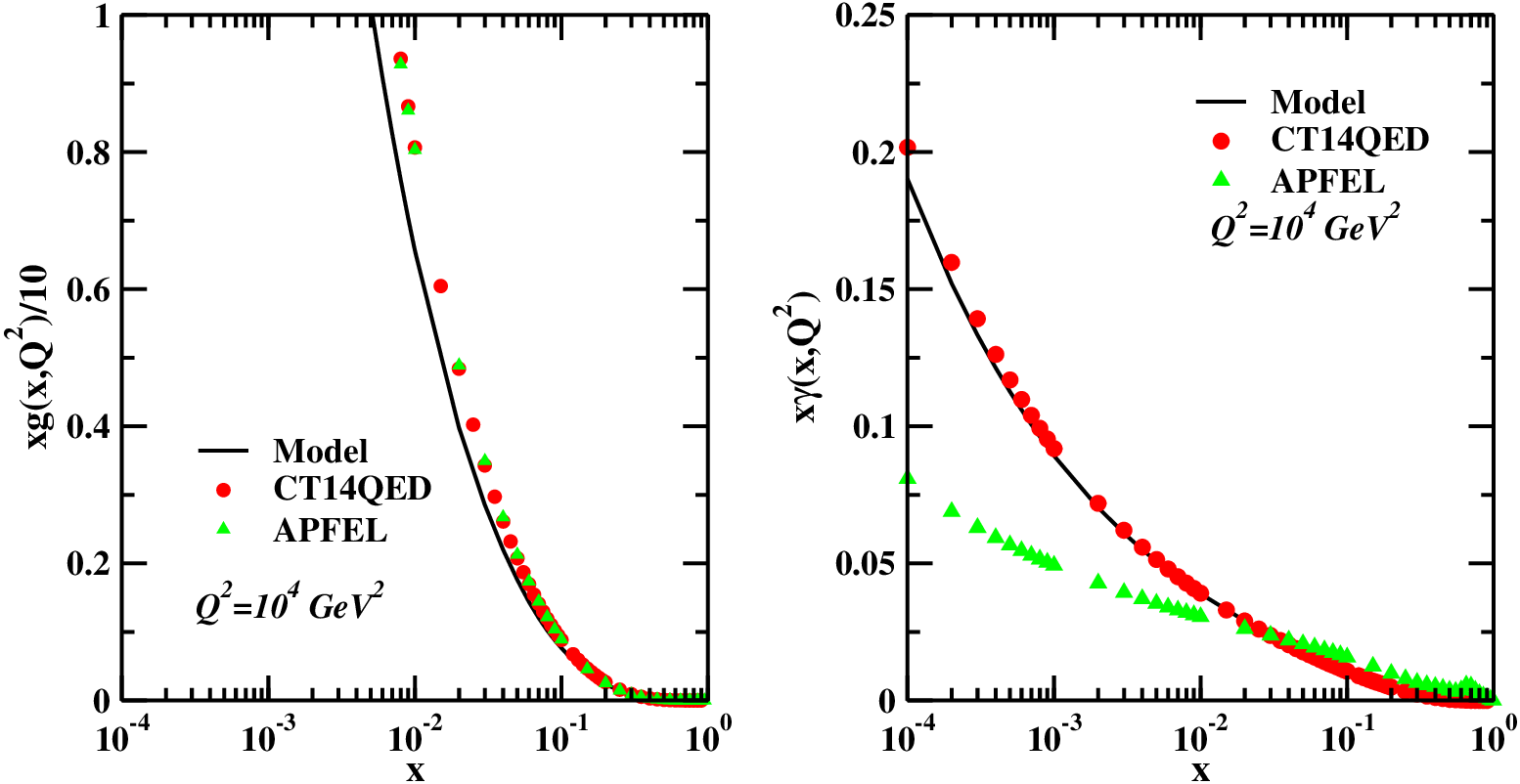}

\caption{The photon and gloun distribution functions at $Q^{2}=10^{4}GeV^{2}$as
a function of x in LO QED and NLO QCD approximations in comparison
with the available APFEL and CT14QED models.}

\label{fig3}
\end{figure}

\begin{figure}[tph]
\centering
\includegraphics[scale=0.5]{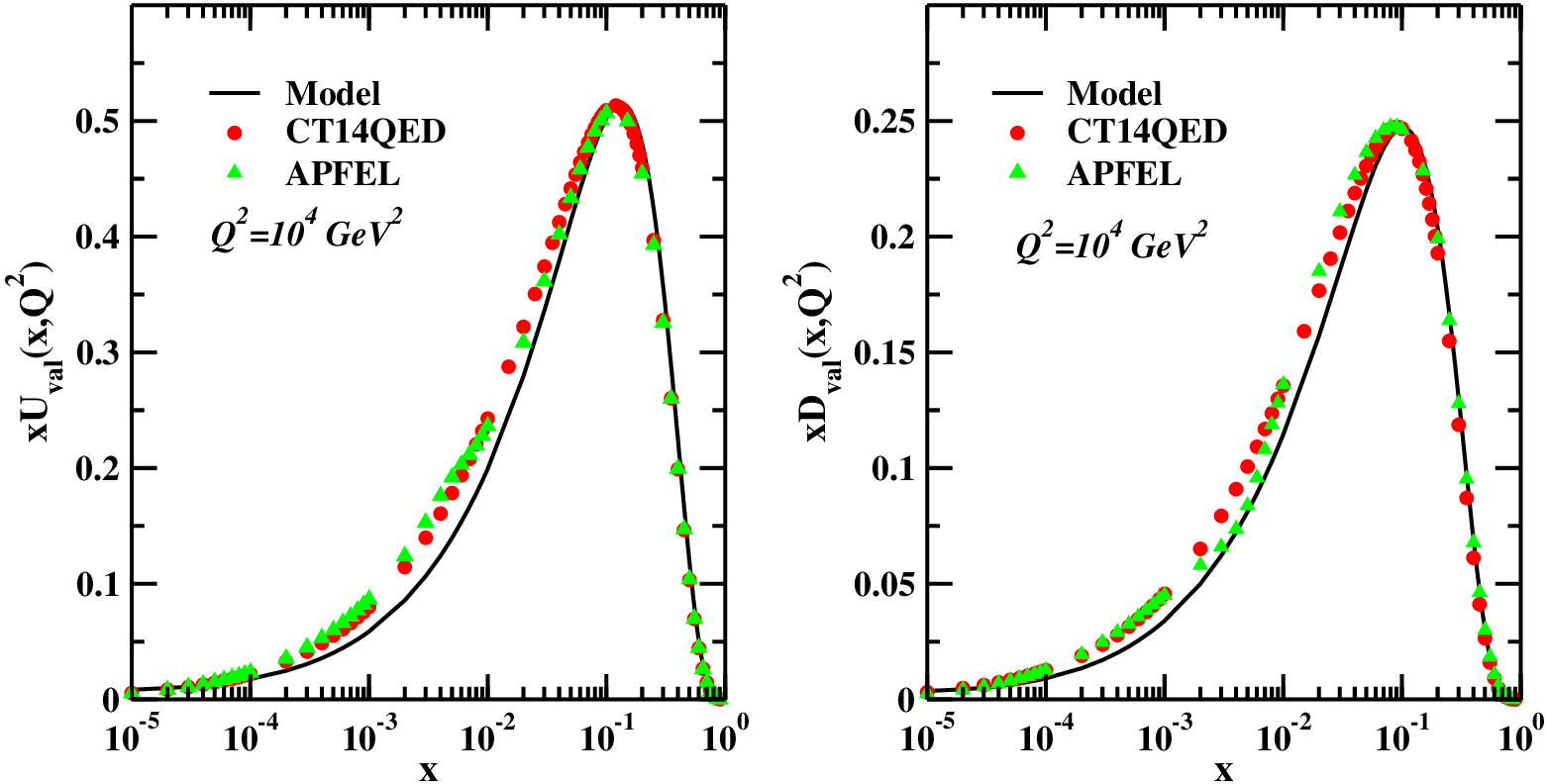}

\caption{The comparison valance quark distributions at $Q^{2}=10^{4}GeV^{2}$
as a function of x with the availabe CT14QED and APFEL models.}

\label{fig4}
\end{figure}

\begin{figure}[tph]
\centering
\includegraphics[scale=0.5]{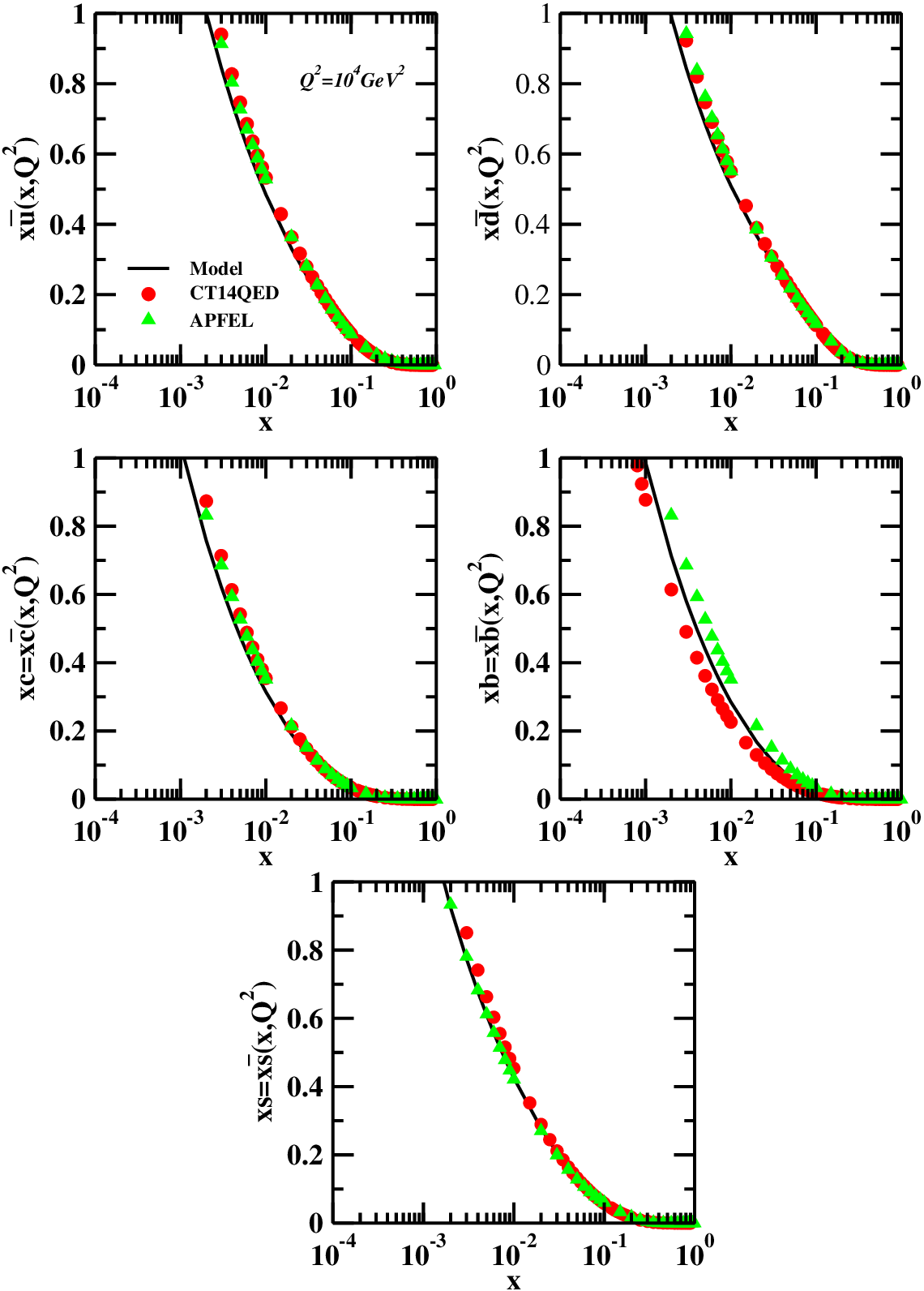}

\caption{The comparison sea quark distributions at $Q^{2}=10^{4}GeV^{2}$as
a function of x in LO QED and NLO QCD approximations with the availabe
CT14QED and APFEL models.}

\centering{}\label{fig5}
\end{figure}

In this section, we will present our results that we are obtained
for the parton distribution functions and proton structure function,
$F_{2}^{p}(x,Q^{2})$, using the Laplace transform technique. The
results are displayed in Figs. ~\ref{fig1}
to ~\ref{fig5}. It should be noted that
we need some initial inputs for PDFs, equations ~\ref{eq17}
and ~\ref{eq22}. We borrowed data for
initial inputs from CT14QED code ~\cite{Schmidt:2015zda}
at $Q_{0}=1.295\,GeV$, to be sure about the correctness of our solutions.
We fit this data with functions in x space and convert these functions
by using Laplace Transforms from x space to s space and then use them
as the initial conditions to getting solutions for DGLAP equations.
These functions are represented in the Table. ~\ref{table1}.
If the solutions are correct then we expect that our PDFs set and
proton structure function have good agreement with those from all
global parameterizations (as well as CT14QED) and experimental data.

\begin{table}[tph]
\caption{The distributions of \textbf{$\mathbf{\mathit{xf}_{i0}}$} as the
initial inputs}

\begin{centering}
\begin{tabular}{c>{\centering}m{12cm}}
\hline
\textbf{$\mathbf{\mathit{xf}_{i0}}$} & \tabularnewline
\hline
$xf_{10}$ & \begin{singlespace}
\centering{}$(-14767.2x^{1.5}+105659.x^{2}-3585.68x-0.960083)(1-x)^{2.91524}/(1+20724.9x)$
\end{singlespace}
\tabularnewline
\hline
$xf_{20}$ & \begin{singlespace}
\centering{}$0.28x^{-0.238}(4.967x^{0.5}+1.27x^{2}+14.98x+1)(1-x)^{3.14}$
\end{singlespace}
\tabularnewline
\hline
$xf_{30}$ & \begin{singlespace}
\centering{}$27.6584x^{0.457605}(1+5.12808x-3.96762x^{0.5}-2.17654x^{2})(1-x)^{5.13677}$
\end{singlespace}
\tabularnewline
\hline
$xf_{40}$ & \begin{singlespace}
\centering{}$0.0135x^{-0.0012}(1-x)^{1.14}(1-2.4x^{0.5}+1.49x)$
\end{singlespace}
\tabularnewline
\hline
$xf_{50}$ & \begin{singlespace}
\centering{}$1.18x^{0.568}\left(1+3.8x-4.78x^{2}\right)(1-x)^{3.73}$
\end{singlespace}
\tabularnewline
\hline
$xf_{50}$ & \begin{singlespace}
\centering{}$1.18x^{0.568}\left(1+3.8x-4.78x^{2}\right)(1-x)^{3.73}$
\end{singlespace}
\tabularnewline
\hline
$xf_{60}$ & \begin{singlespace}
\centering{}$1.79x^{0.55}(1+5.6x)(1-x)^{3.7}$
\end{singlespace}
\tabularnewline
\hline
$xf_{70}$ & \begin{singlespace}
\centering{}$0.0059x^{-0.416}(1+571.1x-1342.33x^{2}+2464.27x^{2.5})(1-x)^{4.83}$
\end{singlespace}
\tabularnewline
\hline 
$xf_{80}$ & \begin{singlespace}
\centering{}$0.156x^{-0.21}(1+20.12x+2.41x^{0.5}+9.57x^{1.5})(1-x)^{3.03}$
\end{singlespace}
\tabularnewline
\hline 
$xf_{90}$ & \begin{singlespace}
\centering{}$0.172x^{-0.184}\left(1+0.0033x^{0.5}\right)(1-x)^{6.23}$
\end{singlespace}
\tabularnewline
\hline
\end{tabular}
\par\end{centering}
\centering{}\label{table1}
\end{table}

The valance quark distributions, $xU_{val}(x,\,Q^{2})$ and $xD_{val}(x,\,Q^{2})$,
at LO QED and NLO QCD approximations are depicted in Figs.~\ref{fig1}
and ~\ref{fig2}. We also compare them with
APFEL model results for the different values of $Q^{2}$. The solid
curves show our results for the valence quark distributions, and the
scatter curves present the APFEL model results. The agreement with
both the d and u valance quark distributions, over the large range
of x and $Q^{2}$, is excellent. The results show that our analytical
solutions for the QED$\otimes$QCD DGLAP evolution equations are
correct and these solutions are correctly used to calculate the parton
distribution functions.

The comparison photon distribution function, $x\gamma(x,Q^{2})$,
gloun distribution function, $xg(x,Q^{2})$,with APFEL and CT14QED
models at $Q^{2}=10^{4}GeV^{2}$ for $\alpha_{s}(Q^{2}=M_{z}^{2})=0.118$
is well demonstrated in Fig.~\ref{fig3}.
This plot indicates that our results are in good agreement with APFEL
and CT14QED models. Also it is clear from this figure for photon distribution
function that our results in comparison with the CT14QED photon distribution
function are very similar at large value of x and are different for
small value of x. We also investigate the effect of an increasing
in value of $Q^{2}>Q_{0}^{2}$ on the photon distribution functions
and conclude that the CT14QED photon distribution function becomes
large, whereas our results is distinctly different and much smaller
at small values of x ( Corresponding plot omitted for briefly).

In figure ~\ref{fig4} we displayed the valance
quark distributions at scale of $Q^{2}=10^{4}GeV^{2}$. we compared
those with the APFEL and CT14QED models. It is shown that with increasing
the value of $Q^{2}$, the contribution of valence quarks are decreased.
Therefore, we can conclude the photon contribution is now significantly
considerable.

Fig. ~\ref{fig5} displays our analytical
sea quark distribution functions at $Q^{2}=10^{4}GeV^{2}$. We compared
our results with the newly released PDFs global parameterizations
from CT14QED ~\cite{Schmidt:2015zda} and APFEL model. The
CT14QED, is the first set of CT14 parton distribution functions obtained
by including QED evolution at leading order with next-to-leading order
QCD evolution in their global analysis.

It is found that the sea quark distribution functions in comparison
with the photon distribution function in the large values of x with
increasing the value of $Q^{2}$, contribution of photon is most significant.
It may also be noted that in range of high x the photon distribution
function is larger than the bottom quark distribution function as
increasing the value of $Q^{2}$.

It is observed from these figures with increasing the value of $Q^{2}$
that the parton distribution functions decrease for the large values
of x and increase for the small values of x. 

We now proceed by calculating proton structure function. Our aim of
to investigate the proton structure function is to compare our results
with a physical observable that confirm the correctness of our analytical
solutions. The Laplace transform technique is also applied to the
proton structure function, $F_{2}^{p}(x,Q^{2})$, which leads to an
analytical solution for this function. The method illustrated in this
analysis enable us to achieve strictly analytical solution for proton
structure function in terms of x variable.

We will yield the total proton structure functions as $F_{2}^{p,total}(x,Q^{2})=F_{2}^{p,light}(x,Q^{2})+F_{2}^{heavy}(x,Q^{2})$
where $F_{2}^{heavy}(x,Q^{2})=F_{2}^{c}(x,Q^{2})+F_{2}^{b}(x,Q^{2})$
are the charm and bottom quarks structure functions. 

For light quarks, the proton structure function $F_{2}^{p,light}(x,Q^{2})$
in Laplace s space, up to the next-to-leading order approximation
is given by
\begin{equation}
F_{2}^{p,light}(s,\tau)=F_{2}^{NS}(s,\tau)+F_{2}^{S}(s,\tau)+F_{2}^{G}(s,\tau),\label{eq:23}
\end{equation}
where the non-singlet $F_{2}^{NS}$, singlet $F_{2}^{S}$ and gloun
$F_{2}^{G}$ contributions are written as

\begin{eqnarray}
& F_{2}^{NS}(s,\tau)=(\frac{4}{9}u_{v}(s,\tau)+\frac{1}{9}d_{v}(s,\tau))(1+\frac{\tau}{2\pi}C_{q}^{(1)}(s))
\nonumber\\&
F_{2}^{S}(s,\tau)=(\frac{4}{9}2\bar{u}(s,\tau)+\frac{1}{9}2\bar{d}(s,\tau)+\frac{1}{9}2\bar{s}(s,\tau))
(1+\frac{\tau}{2\pi}C_{q}^{(1)}(s))
\nonumber\\&
F_{2}^{G}s,\tau)=(\frac{4}{9}+\frac{1}{9}+\frac{1}{9})g(s,\tau)(\frac{\tau}{2\pi}C_{g}^{(1)}(s))\label{eq:25}
\end{eqnarray}

where the $C_{q}^{(1)}(s)$ and $C_{g}^{(1)}(s)$ are the next-to-leading
order Wilson coefficient functions, derived in Laplace s space by
$C_{q}(s)=L[e^{\text{\textminus}\nu}c_{q}(e^{\text{\textminus}\nu});s]$
and $C_{g}(s)=L[e^{-\nu}c_{g}(e^{-\nu});s]$. The next-to-leading
order Wilson coefficient functions in Bjorken x space are found in
refs. ~\cite{Bardeen:1978yd}. We have found the
final desired solution of the proton structure function in x space,
$F_{2}^{p,light}(x,Q^{2})$, using the inverse Laplace transform and
the appropriate change of variables.

The next-to-leading order contribution of heavy quarks, $F_{2}^{c,b}(x,Q^{2})$,
to the proton structure function can be calculated in the fixed flavour
number scheme (FFNS) approach ~\cite{Gluck:2007ck,Gluck:2004fi,Gluck:2008gs,Gluck:1993dpa,Laenen:1992cc,Riemersma:1994hv,Laenen:1992zk}. 

The heavy quarks structure function, $F_{2}^{c,b}(x,Q^{2})=F_{2}^{(nl)}(x,Q^{2})+F_{2}^{(d)}(x,Q^{2})$,
where $F_{2}^{(nl)}(x,Q^{2})$ and $F_{2}^{(d)}(x,Q^{2})$ are the
massive-scheme heavy-quark structure function and the \textquotedblleft difference\textquotedblright{}
contribution, respectively. The Laplace Transforms of $F_{2}^{(nl)}(x,Q^{2})$
and $F_{2}^{(d)}(x,Q^{2})$ for charm and bottom quarks, are given
by
\begin{equation}
F_{2}^{(nl)}(s,\tau)=\frac{4}{9}\tau
\left(C_{g}^{(1)}(s)\:Log\left(\frac{Q^{2}}{m_{c}^{2}}\right)+C_{g}^{(1)}(s)\right)g(s,\tau)\label{eq:26}
\end{equation}
\begin{equation}
F_{2}^{(d)}(s,\tau)=\frac{4}{9}\left(1+\frac{\tau}{2\pi}C_{q}^{(1)}(s)\right)(c(s,\tau)+\bar{c}(s,\tau))
+\frac{4}{9}\frac{\tau}{2\pi}(C_{g}^{(1)}(s)-C_{g}^{(1)}(s,m_{c}^{2}))\,g(s,\tau)\label{eq:27}
\end{equation}
and 
\begin{equation}
F_{2}^{(nl)}(s,\tau)=\frac{1}{9}\tau\\
\left(C_{g}^{(1)}(s)\:Log\left(\frac{Q^{2}}{m_{b}^{2}}\right)+C_{g}^{(1)}(s)\right)g(s,\tau)\label{eq:28}
\end{equation}
\begin{equation}
F_{2}^{(d)}(s,\tau)=\frac{1}{9}\left(1+\frac{\tau}{2\pi}C_{q}^{(1)}(s)\right)(b(s,\tau)+\bar{b}(s,\tau))\\
+\frac{1}{9}\frac{\tau}{2\pi}(C_{g}^{(1)}(s)-C_{g}^{(1)}(s,m_{b}^{2}))\,g(s,\tau)\label{eq:29}
\end{equation}
where $m_{c}$ and $m_{b}$ are the charm and bottom quark masses.
The coefficient functions $C_{g}^{(1)}(s,m_{c}^{2})$ and $C_{g}^{(1)}(s,m_{b}^{2})$
are found in ref.~\cite{Forte:2010ta}.

\begin{figure}[tph]
\centering
\includegraphics[scale=0.5]{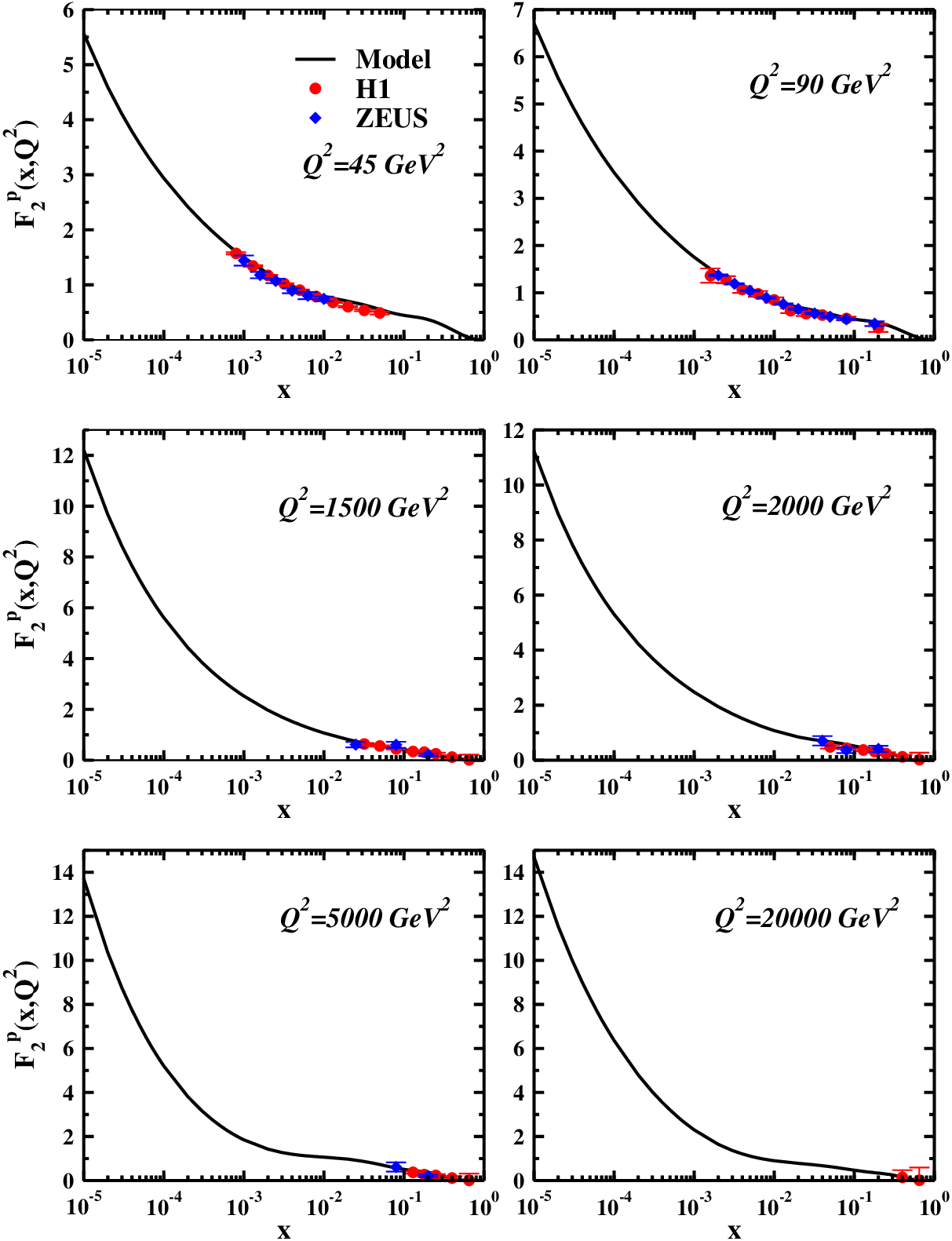}
\caption{The proton structure function at $Q^{2}=45,\,90,\,1500,\,2000\,,5000\:and\:20000\:GeV^{2}$
in comparison with experimental data.}
\label{fig6}
\end{figure}

\begin{figure}[tph]
\centering
\includegraphics[scale=0.5]{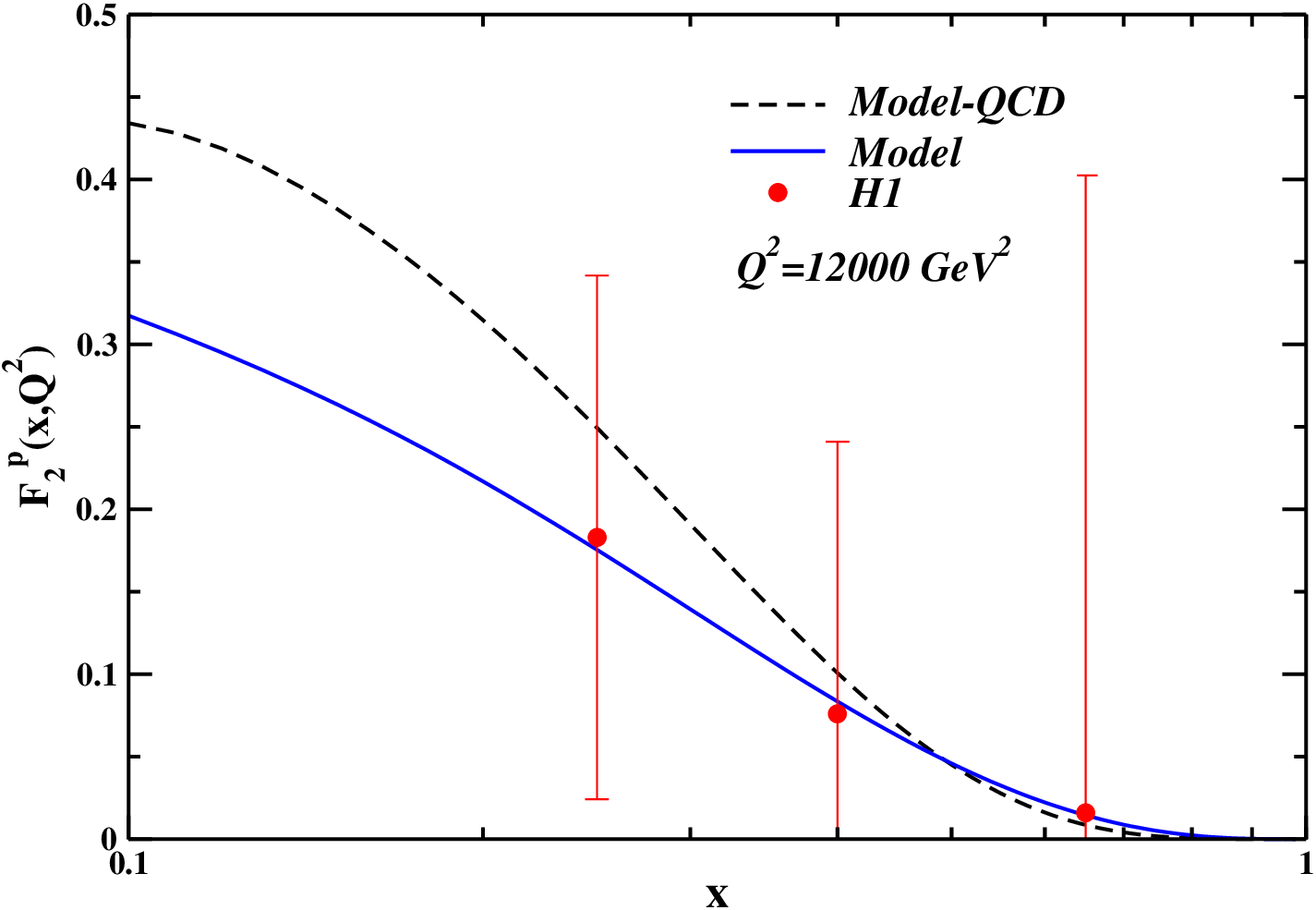}

\caption{The proton structure function at $Q^{2}=12000GeV^{2}$in comparison
with QCD analysis and experimental data }

\label{fig7}
\end{figure}

Figure~\ref{fig6} depicts the
comparison the proton structure function with the corresponding available
experimental data from the H1 and ZEUS Collaborations in the several
values of $Q^{2}$. The results demonstrate that there are good agreement
between them. It is clear that the proton structure function increases
with an increase in value of $Q^{2}$ for small values of x and decrease
for large values of x. All figures indicate that the analytical solutions
work well beyond the charm quark mass threshold, $Q^{2}>Q_{0}^{2}(\approx m_{c}^{2}=1.677\,GeV^{2})$. Figure~\ref{fig7} displays the comparison the proton structure function with QED corrections and
without this corrections (QCD analysis) with the corresponding experimental
data from the H1 Collaborations in the value of $Q^{2}=12000\,GeV^{2}$.
This figure shows that the proton structure function with QED corrections
in good agreement with experimental data in the high energy.

\section{CONCLUSIONS}\label{sec:con}

In this paper, we utilized the Laplace transform technique to calculate the Laplace transformation of splitting functions and extract the parton distribution functions of quark, antiquark, gloun and photon inside the proton. Our calculations are done in LO QED and NLO QCD approximations. We finally extracted the unpolarized proton structure
functions at the different values of $Q^{2}$. Our results are compared with APFEL and newly released CT14QED codes and also with experimental data which indicate good agreements between them. To determine the proton structure function at any arbitrary $Q^{2}$ scale , we only need to know the initial distributions for singlet, gluon, non-singlet
and photon distributions at the input scale of $Q_{0}^{2}$. We borrowed the initial inputs from CT14QED code at $Q_{0}=1.295\,GeV$ to be sure about the correctness of our solutions. The solutions are seem to be correct because the parton distribution functions and the proton
structure function have good agreement with those from all global parameterizations (as well as CT14QED) and experimental data. In the future work with a global parameterization we can determine these initial inputs. These PDFs can specifically design for use in precision cross section predictions and uncertainties at the LHC.

\begin{center}
\textbf{ACKNOWLEDGMENT}
\par\end{center}

We would like to thank Professor S. Atashbar Tehrani for his help
and for the productive discussions.

\appendix
\begin{center}
\textbf{APPENDIX: MATHEMATICA PROGRAM OF THE SPLITTING FUNCTIONS}
\par\end{center}

Program containing our results for the Laplace transforms of the splitting
functions at LO QED and NLO QCD approximations can be obtained via
Email from the authors upon request.

\end{document}